\documentclass[%
 reprint,
 amsmath,amssymb,
 aps,
 prapplied,
]{revtex4-2}

\usepackage{graphicx}
\usepackage{dcolumn}
\usepackage{bm}
\usepackage{hyperref}
\usepackage{xr}
\usepackage{xcolor}

\newcommand{\red}[1]{\textcolor{black}{#1}}
\newcommand{\redd}[1]{\textcolor{black}{#1}}

\begin{document}

\title{Efficient Radiofrequency Sensing with Fluorescence Encoding}

\author{N. Voce}
\affiliation{Department of Physics, Northeastern University}
\author{P. Stevenson}
\affiliation{Department of Physics, Northeastern University}

\date{\today}

\begin{abstract}
Optically-active spin qubits have emerged as powerful quantum sensors capable of nanoscale magnetometry, yet conventional coherent sensing approaches are ultimately limited by the coherence time of the sensor, typically precluding detection in the sub-MHz regime. We present a broadly applicable fluorescence-encoding method that circumvents coherence-time constraints by transducing time-varying magnetic fields directly into modulated fluorescence signals. Using nitrogen-vacancy centers in diamond as a model system, we demonstrate shot-noise-limited sensitivity for AC magnetic fields spanning near-DC to MHz frequencies, with detection bandwidth tunable via optical excitation power. The technique captures complete spectral information in a single measurement, eliminating the need for point-by-point frequency scanning, and allows phase-sensitive multi-frequency detection. This approach transforms quantum sensors into atomic-scale spectrum analyzers, with immediate applications for low-frequency RF communication, zero-field NMR, and bioelectronic sensing. Our approach is broadly applicable to the expanding class of optically-active spin qubits, including molecular systems and fluorescent proteins, opening new sensing regimes previously inaccessible to coherent techniques.
\end{abstract}

\maketitle

\section{Introduction}

Solid-state quantum sensors - specifically, optically-active spin qubits - have emerged as outstanding magnetometers with excellent sensitivity, nanoscale spatial resolution, and a broad range of deployable form-factors, spanning single sensors to macroscopic ensembles~\cite{chatterjeeSemiconductorQubitsPractice2021,schirhaglNitrogenVacancyCentersDiamond2014,casolaProbingCondensedMatter2018}. This combination of properties has enabled unique experimental capabilities that range from single-molecule nuclear magnetic resonance (NMR) experiments~\cite{mullerNuclearMagneticResonance2014,lovchinskyNuclearMagneticResonance2016} to imaging current flow in low-dimensional systems~\cite{andersen_electron-phonon_2019,palm_imaging_2022,ku_imaging_2020,PhysRevLett.129.087701}.

Existing sensing approaches, however, face two related challenges. First, coherent sensing of time-varying (AC) signals is limited by sensor coherence time ($T_2$): sensitivity degrades dramatically for frequencies below $1/T_2$, even when novel correlation methods are used~\cite{laraouiDiamondNitrogenvacancyCenter2011,bossQuantumSensingArbitrary2017,schmittSubmillihertzMagneticSpectroscopy2017}. Practically, this limits many common sensors to the MHz-and-above range, particularly at room temperature and/or near interfaces. Moreover, emerging classes of quantum sensors - small molecules~\cite{mann_chemically_2025,mena_room-temperature_2024,singh_room-temperature_2025} and fluorescent proteins~\cite{feder_fluorescent-protein_2025,abrahams_quantum_2025,meng_optically_2025} - offer the prospect of integrating sensors directly in chemical and biological systems, but have coherence times significantly shorter than solid-state systems, motivating the need for new sensing strategies. Developing new approaches for frequencies below the $1/T_2$ limit would open up new frontiers in RF communication~\cite{magalettiQuantumRadioFrequency2022,chenQuantumEnhancedRadio2023,hermannExtendingRadiowaveFrequency2024} (enabling direct detection of the low-frequency communication bands), zero-field NMR~\cite{staudacherNuclearMagneticResonance2013,maminNanoscaleNuclearMagnetic2013,steinertMagneticSpinImaging2013}, and bioelectronics~\cite{barryOpticalMagneticDetection2016,webbDetectionBiologicalSignals2021}.

The second challenge lies in measuring the spectrum of time-dependent signals. The power spectral density (PSD) of magnetic noise encodes a vast amount of information and lies at the heart of many proposed applications for quantum sensors. However, existing measurement protocols detect the spectral density at a single frequency~\cite{mazeNanoscaleMagneticSensing2008,balasubramanianUltralongSpinCoherence2009,naydenovDynamicalDecouplingSingleelectron2011,delangeSingleSpinMagnetometryMultipulse2011,kotlerSingleionQuantumLockin2011,ajoyQuantumInterpolationHighresolution2017,wangNanoscaleVectorAC2021} (or, in some cases, over a limited bandwidth~\cite{schmittSubmillihertzMagneticSpectroscopy2017}), which requires many measurements to reconstruct the entire spectrum. This problem is particularly acute in the case of unknown, congested spectra where many hundreds or thousands of individual measurements may be required.

\begin{figure*}
\includegraphics[width=0.9\textwidth]{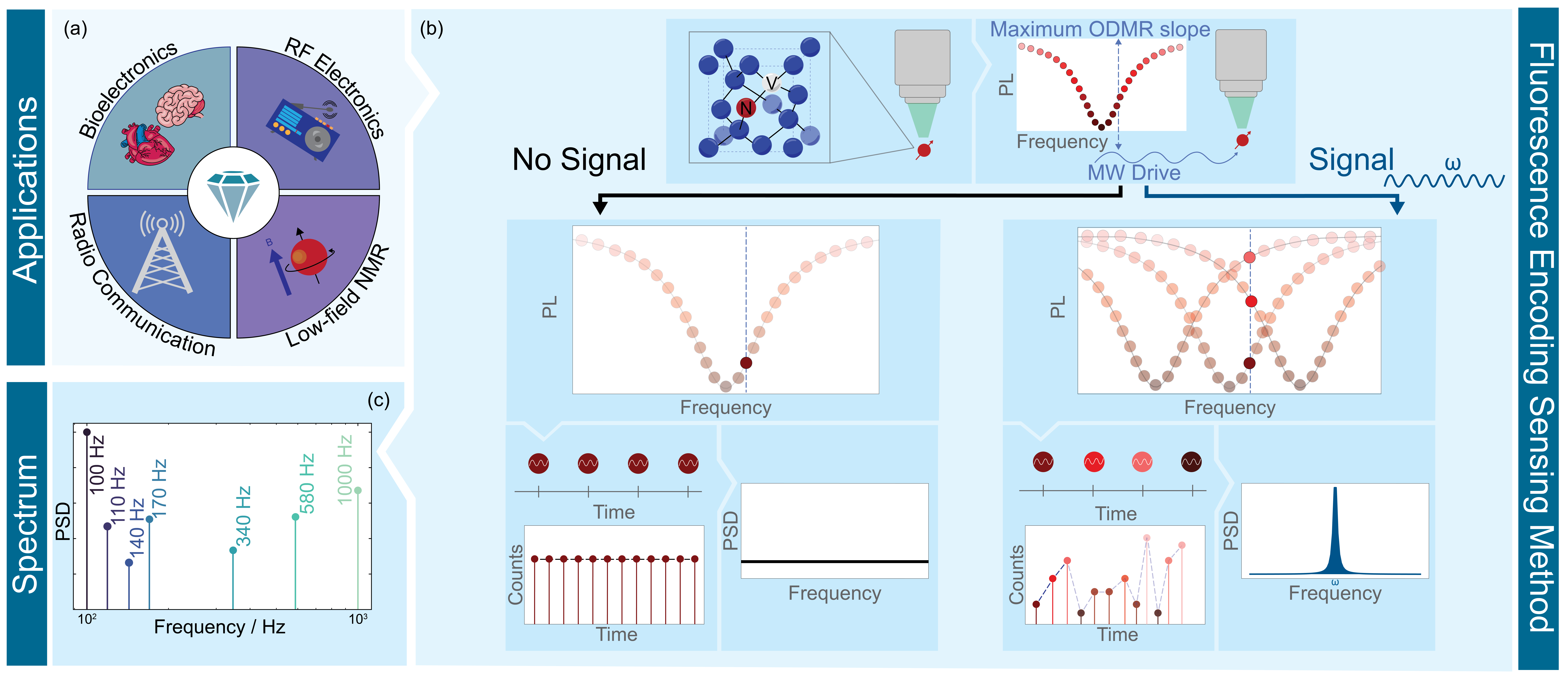}
\caption{(a) Detection of RF signals in the DC-MHz band is essential for a broad range of applications, from bioelectronics to radio communication. (b) The concept of the sensing approach uses the time-dependent frequency shift induced by a target signal to change the fluorescence of a defect such as the nitrogen vacancy center in diamond. Fourier transforming the photon count time series recovers the target signal. \redd{``PL'' refers to photoluminescence.} (c) Example signals, demonstrating both excellent resolution and bandwidth.}
\label{fig:fig1}
\end{figure*}

\red{To address these challenges, we build on previous demonstrations of near-DC sensing\cite{shao2016diamond, barryOpticalMagneticDetection2016} to develop an alternative sensing method based on encoding the target signal onto the fluorescent output of the quantum sensor.} With this approach, we are not only able to sense magnetic fields in the DC-MHz range but also capture rich spectral information in a single experiment, transforming our quantum sensor into an atomic-scale spectrum analyzer.

\section{Results \& Discussion}

\subsection{Operating Principle \& Sensitivity}

To avoid the challenges associated with coherent sensing at low frequencies, namely the necessity of long coherence times, we adopt an alternative, incoherent, fluorescence-encoding scheme, summarized in Fig.~\ref{fig:fig1}. As a model system, we consider the nitrogen vacancy (NV) center in diamond, a fluorescent, paramagnetic ($S=1$) color center, though we emphasize that our approach can be utilized by any of the rapidly growing class of optically-active spin qubits, which has expanded to encompass low-dimensional materials, molecules~\cite{mann_chemically_2025,mena_room-temperature_2024,singh_room-temperature_2025}, and even proteins~\cite{feder_fluorescent-protein_2025,abrahams_quantum_2025,meng_optically_2025}. In the NV center, the fluorescence depends on which spin sublevel is occupied, enabling optical readout of the spin state. Combined with preferential relaxation into the $m_s = 0$ state under optical excitation (generating spin polarization) and microwave control of the spin state, optically-detected magnetic resonance (ODMR) measurements can be performed: sweeping the microwave frequency reveals the spectrum of the NV center spin \emph{via} changes in the fluorescence (Fig.~\ref{fig:fig1}b).

ODMR spectroscopy has been used to great effect to detect and image static magnetic fields; as the magnetic field changes, so do the relative energies of the NV center spin states, changing the position of the ODMR peak. Our approach builds on this phenomenon: by monitoring the fluorescent response under monochromatic microwave excitation, \red{time-dependent changes in magnetic field (changes in the peak position) are transduced into a time-dependent fluorescence signal, previously utilized to directly detect time-domain radio signals\cite{shao2016diamond}.} Here, the Fourier transform of this signal reveals the underlying spectral response across our detection bandwidth (discussed later). For ease of comparison with previous works, we present and discuss our data as the power spectral density (PSD), defined as $S(\omega) = |FT\{F(t)\}|^2$ where $F(t)$ is the time-dependent fluorescence.

\begin{figure}
\includegraphics[width=0.48\textwidth]{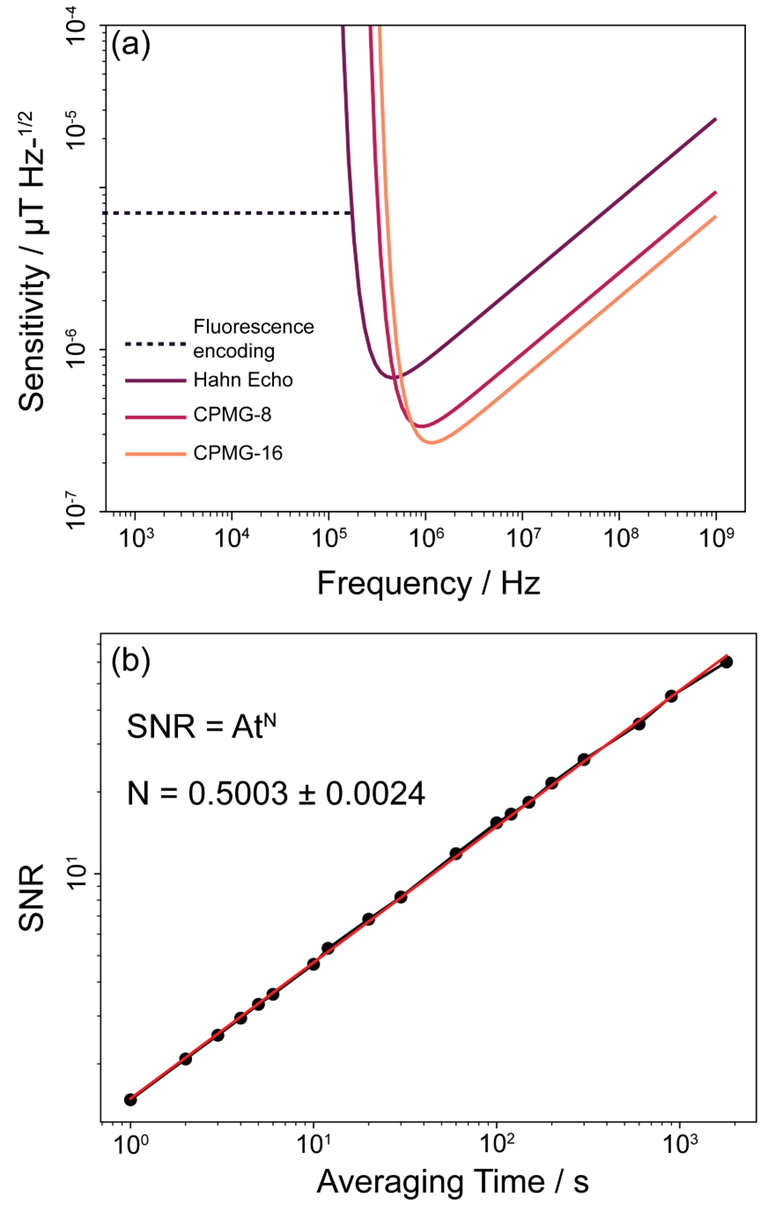}
\caption{(a) Theoretical frequency-dependent sensitivities of Sample A for different sensing techniques. Below 1MHz the fluorescence encoding approach is more sensitive than coherence-based approaches. \red{Sensitivity curves use values obtained from Sample A: $C = 0.1$; $\zeta = 0.09$; $T_2 = 2 \times 10^{-6} s$. (b) Scaling of the empirical signal-to-noise ratio with averaging time for the FE method, showing the $\sqrt{t}$ scaling expected for a shot-noise limited process.}}
\label{fig:fig2}
\end{figure}

Seminal work in detecting individual frequency components with NV centers employed spin-echo pulse sequences~\cite{mazeNanoscaleMagneticSensing2008,balasubramanianUltralongSpinCoherence2009}; since then, dynamical decoupling~\cite{mullerNuclearMagneticResonance2014,staudacherNuclearMagneticResonance2013,kotlerSingleionQuantumLockin2011}, quantum memory~\cite{pfenderNonvolatileNuclearSpin2017,rosskopfQuantumSpectrumAnalyzer2017,zaiserEnhancingQuantumSensing2016}, and heterodyne~\cite{bossQuantumSensingArbitrary2017,schmittSubmillihertzMagneticSpectroscopy2017} approaches have greatly improved spectral resolution, yet have been focused primarily on the MHz spectral range. We show the capability of our fluorescence encoding method in detecting monochromatic signals in the DC-MHz range with a single NV center (Figure~\ref{fig:fig1}c, ensemble data shown in Figure S2a). The resolution of these spectra is limited by the acquisition time (and ultimately clock stability), rather than by the coherence time of the spin qubit. The sensitivity of this approach is maximized when the microwave drive frequency is set to the maximum slope of the ODMR. The shot-noise-limited, single-frequency sensitivity of our sensing approach is equivalent to the well-established DC sensitivity for optically-active spin qubits~\cite{dreau_avoiding_2011} (see Supplemental Information\cite{supp}). For a Lorentzian lineshape, this is:
\begin{equation}
\eta_{FE} = \frac{4}{3\sqrt{3}}\frac{\Gamma}{\gamma}\frac{1}{C\sqrt{R}}
\label{eq:sensitivity}
\end{equation}
where $\Gamma$ is the full-width half-max of the ODMR spectrum, $C$ is the contrast, $R$ is the detection photon rate, and $\gamma$ is the NV center gyromagnetic ratio. This sensitivity is calculated assuming operation at the point of maximum sensitivity, \emph{i.e.,} the steepest part of the Lorentzian curve, $f_{sens} = f_0 + \frac{1}{2}\Gamma/\sqrt{3}$.

For our continuous excitation approach, there is an experimental tradeoff between $\Gamma$, $C$, and $R$: at higher laser powers the detected photon rate ($R$) increases while the ODMR peak broadens ($\Gamma$ increases) and contrast ($C$) decreases~\cite{dreau_avoiding_2011}. Similarly, larger microwave drive amplitudes can increase $C$ but lead to power-broadening effects (larger $\Gamma$). \red{Figure~\ref{fig:fig2}a compares the theoretical shot-noise-limited, frequency-dependent sensitivity for the fluorescence encoding method with conventional Hahn echo and dynamical decoupling-based approaches. To enable robust relative comparisons between these methods, we calculate sensitivities across approaches for the same sample, namely, our nanopillar sample\cite{supp}. Sensitivity was calculated using the following equation \cite{delangeSingleSpinMagnetometryMultipulse2011}:
\begin{equation}
    \eta_{AC}=\frac{1}{\gamma C} \sqrt{\frac{f}{2 N \zeta}}e^{\frac{N}{(2fT_2)^3}}
    \label{eq:AC_sen}
\end{equation}
where $\gamma$ is the gyromagnetic ratio, $C$ is the contrast, $f$ is the signal frequency, $N$ is the number of $\pi$ pulses in the pulse sequence, $\zeta$ is the number of photons collected per measurement, and $T_2$ is the coherence time. For Figure~\ref{fig:fig2}a we use the following experimentally-determined parameters: $C = 0.1$; $\zeta = 0.09$; $T_2 = 2 \times 10^{-6} s$. 
The data in Figure~\ref{fig:fig2}a illustrates the complementary nature of these approaches: the sub-MHz regime, where coherence constraints are onerous, is well-served by the fluorescence encoding strategy.}

\begin{table}
\caption{\red{Summary of NV center sample parameters and ideal sensitivities used in this work. For single-center measurements, reported values are for 30~$\mu$W incident laser power, measured after the objective. The beam waist at the focus is $\approx300$\,nm.} This is below the saturation power to avoid excess line broadening.}
\label{tab:samples}
\begin{ruledtabular}
\begin{tabular}{lcccc}
Sample & Linewidth & Contrast & Count Rate & $\eta_{ideal}$ \\
 & (MHz) & (\%) & ($s^{-1}$) & ($\mu$T/$\sqrt{\text{Hz}}$) \\
\hline
A (NV15) & 9.6 & 11.62 & 72,000 & 8.5 \\
A (NV32) & 8.0 & 10.58 & 60,000 & 8.5 \\
B (Ensemble) & 15 & 1.61 & 3,700,000 & 13.3 \\
\end{tabular}
\end{ruledtabular}
\end{table}

For experimental demonstrations, we use two very different samples: Sample A, a nanopillar structure hosting single NV centers, and Sample B, an ensemble of NV centers formed in high-pressure-high-temperature synthesized diamond. For both samples, we calculate similar shot-noise-limited sensitivities (Table~\ref{tab:samples}); the higher count rate for the HPHT sample is offset by its smaller contrast, though we note that we deliberately attenuate our fluorescence detection rate for the HPHT sample to avoid saturating the single-photon detector.

Figure~\ref{fig:fig2}b shows the measured signal-to-noise ratio (SNR) for a calibrated test signal of $4.5~\mu$T for a single NV center in Sample A (see Supplemental Information \cite{supp} for more details on the calibration procedure). Fitting this data to the form $SNR = At^b$ yields $A = 1.4$ and $b = 0.5$, demonstrating the $\sqrt{N}$ scaling ($N$ is the total photons collected) we expect for a shot-noise limited process. The empirical sensitivity we calculate from this is $3.2~\mu$T/$\sqrt{\text{Hz}}$; at first glance, this is better than expected for the shot-noise limited case (Table~\ref{tab:samples}). However, in the limit of low laser and microwave powers, the lineshape of the NV center is not well-described by a single Lorentzian (as assumed in the previous analysis) and instead is a combination of peaks separated by the hyperfine splitting; in our case ($^{14}$N isotopes) this yields three peaks separated by $2.1$ MHz~\cite{dreau_avoiding_2011}. Fitting to this more correct functional form yields the true peak linewidth (4.04 MHz vs 8.0 MHz for the single peak fit \cite{supp}), which brings the estimated sensitivity into better agreement with the measurements. For simplicity, however, we adopt the more conservative single-Lorentzian fit peak width when calculating and discussing sensitivities throughout this work.

These sensitivity estimates, however, assume that a change in NV center frequency translates to a linear shift in fluorescence; this is a good assumption only for small-amplitude test signals. This requires that $\Delta \ll \sqrt{3}\Gamma$ \cite{supp}, where $\Delta$ is the magnitude of the frequency shift caused by the target signal. For a representative linewidth of 8 MHz, this translates to a requirement that our test field peak-to-peak value should be $\ll 500~\mu$T. Signals beyond this limit manifest as nonlinear responses to the target signal, resulting in additional peaks at harmonics of the signal frequency (Figure S6).

\subsection{Detection Bandwidth}

\begin{figure}
\includegraphics[width=0.48\textwidth]{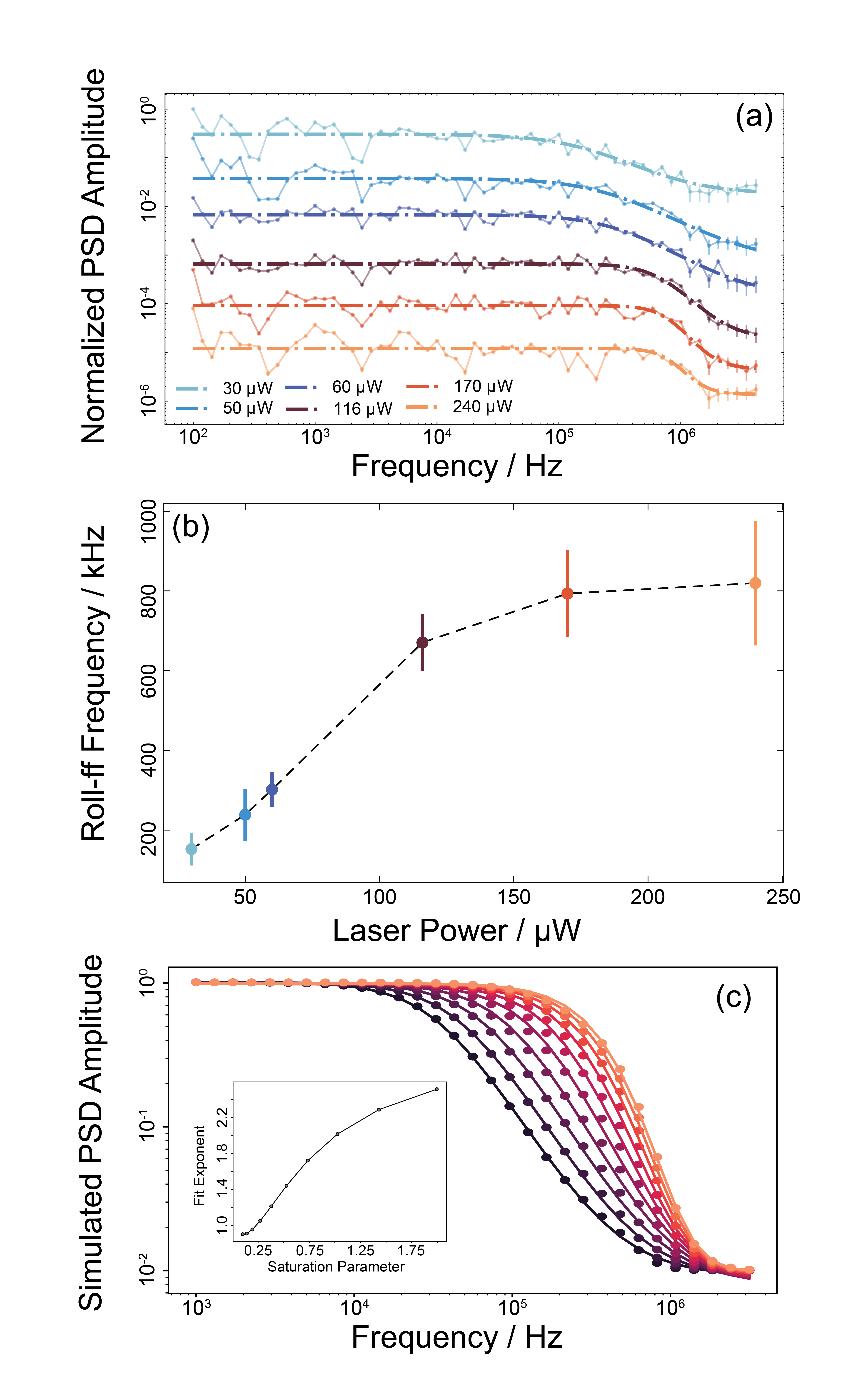}
\caption{(a) Frequency response of the NV center as a function of excitation laser power. Curves are offset for visual clarity (b) Bandwidth as a function of laser power, showing the increase with laser power. (c) Simulated response curves as a function of saturation parameter from low (black) to high (orange). Solid lines are fits to the empirical function described in the text. \red{Inset shows the evolution of the exponent as a function of saturation parameter $s=k_{ex}/k_{em}$, consistent with the experimental data.}}
\label{fig:fig3}
\end{figure}

In our fluorescent encoding method, the time-varying magnetic field is transduced to an optical signal \emph{via} the spin-dependent fluorescence. The question arises: what physical limits are there on the sensing bandwidth of our approach? We expect that signals changing much faster than the average time between emitted photons will not be detected; given the lifetime of NV centers (13-18 ns depending on environment) this suggests a potential upper limit on bandwidth of $\sim$50 MHz. However, the NV center photocycle also involves singlet shelving states with lifetimes on the order of $>$500 ns, whose steady-state population under illumination is intensity-dependent~\cite{tetienneMagneticfielddependentPhotodynamicsSingle2012}. Together, this suggests the detection bandwidth is dependent on optical excitation power; the important metric is the average time between emitted photons. In the high optical power limit, this is determined by a combination of the excited state lifetime and shelving state, while in the low power limit, this is instead limited by the excitation rate. Figure~\ref{fig:fig3}a (Sample A) shows the response of the NV center to test signals of varying frequency, as a function of optical excitation power, \red{complementing previous work characterizing detection bandwidth scaling with microwave power\cite{shao2016diamond}.} As we expect, the upper frequency limit increases as the excitation power increases, consistent with previous observations of ensemble NV centers probed with a dissipative-quantum-sensing protocol~\cite{xieDissipativeQuantumSensing2020}.

To evaluate the cutoff frequency as a function of laser power, we fit the observed response to the functional form:
\begin{equation}
S(f) = \frac{A}{\left(1 + \left(\frac{f}{f_c}\right)^2\right)^b} + c
\label{eq:bandwidth}
\end{equation}
where $S(f)$ is the PSD, $A$ is the amplitude, $f$ is the test signal frequency, $f_c$ is the bandwidth, or cutoff frequency, $b$ is the power-law exponent controlling the roll-off steepness, and $c$ is a constant offset. This functional form is empirically chosen because it captures the evolution of the response curve with laser power, allowing us to extract the cutoff frequency. In the case $b = 1$, this is the expected PSD for a low-pass RC filter, though we regard $b$ only as an empirical parameter describing changes in the functional form.

We sweep laser power and observe the expected trend of cutoff frequency increasing with laser power (Fig.~\ref{fig:fig3}b). We note that the bandwidth reported here is not a hard cut-off; we are still able to detect attenuated responses beyond these frequencies. The frequency dependence we measure has some fine structure, observable as reproducible dips in the response; we attribute this to frequency-dependent transmission in the RF excitation line in our experiment. Optimizing the cables and electronics used removes these features (\cite{supp}).

To further investigate the interplay of optical and microwave driving with the spin dynamics, we perform numerical simulations of the Lindblad master equation. For a set of experimentally informed parameters, we are able to recover not only the approximate bandwidth of our system, but we also observe a change in the exponent of Eq.~\ref{eq:bandwidth} (Fig.~\ref{fig:fig3}c, inset), as in our experiment (Figure S8). As the laser excitation rate increases, the effective spin-lattice relaxation time (\emph{i.e.,} the average time we expect a spin to stay in a particular population state) decreases. \red{To enable comparison between simulation and experiment, we define the saturation parameter $s$, which is defined in the simulations as the ratio of the excitation rate to the spontaneous emission rate: $s=k_{ex}/k_{em}$. }This increases the effective bandwidth but comes at the cost of sensitivity; the fast optical pumping broadens the NV center linewidth, decreasing the sensitivity.

\subsection{Multi-frequency detection}

\begin{figure}
\includegraphics[width=0.48\textwidth]{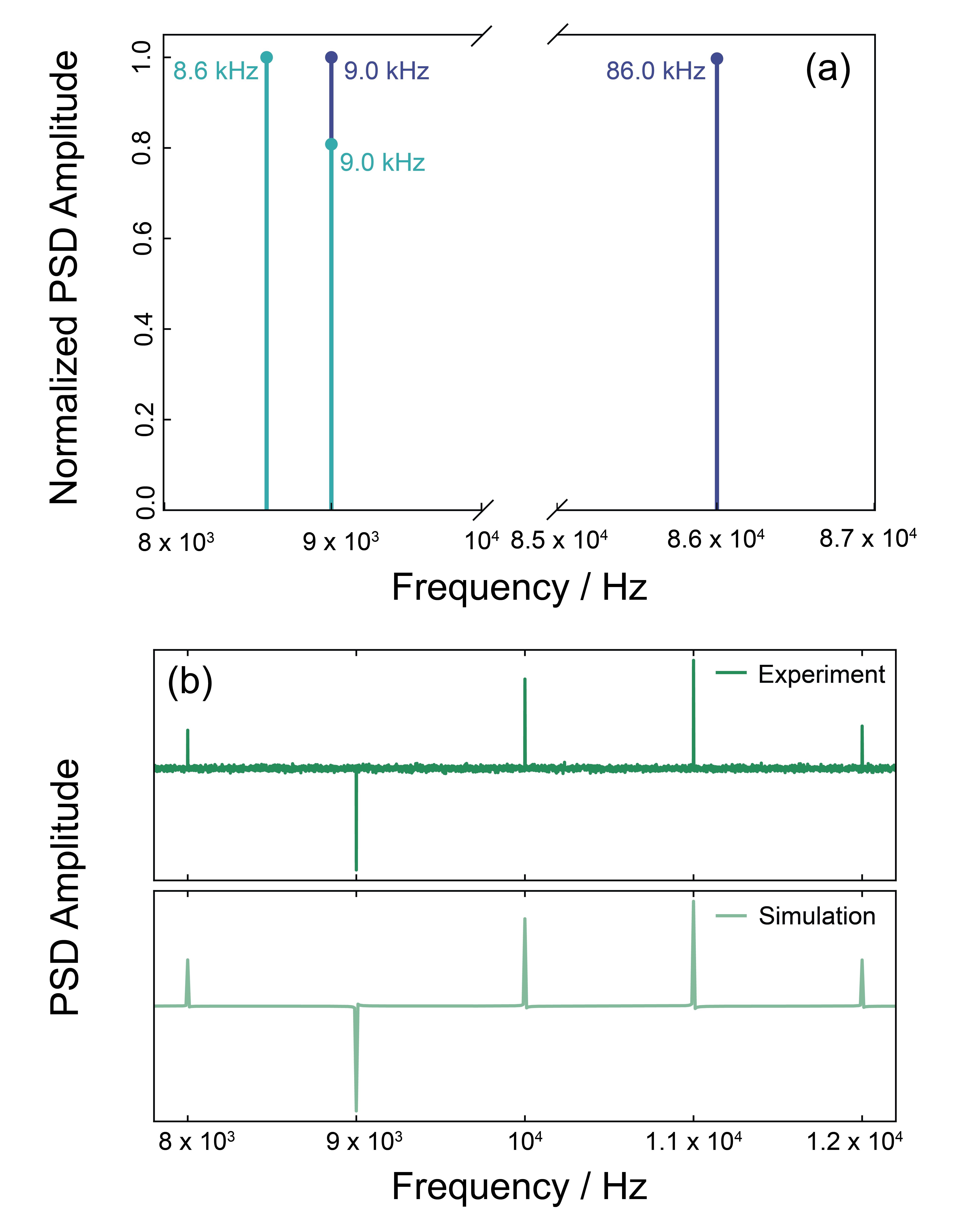}
\caption{Multifrequency detection: (a) the fluorescence encoding approach can simultaneously detect signals close or far in frequency space. (b) With appropriate reference signal, phase sensitive detection can be realized. Here, a phase modulated signal is recovered and compared against the expected analytical form.}
\label{fig:fig4}
\end{figure}

Sparse, unknown spectra pose a challenge for point-by-point sensing methods, where, unless the spectral structure is known \emph{a priori}, most of the measurement time is spent measuring frequencies where there is no response~\cite{schmittSubmillihertzMagneticSpectroscopy2017}. Thus, the ability to detect multiple frequencies simultaneously represents a significant advantage in frequency-domain sensing; Fig.~\ref{fig:fig4}a demonstrates the simultaneous detection of multiple sine waves with both closely spaced frequencies (8.6 kHz vs 9 kHz) and far-detuned frequencies (9 kHz and 86 kHz) for Sample A. (Equivalent data for Sample B is shown in Figure S2b.) The spectral range of our technique is limited only by acquisition time and/or clock stability (frequency resolution) and detection bandwidth.

We make a distinction here when discussing sensitivity between the single-point sensitivity (\emph{i.e.,} how quickly one can determine the amplitude of a signal at a known frequency) and the true acquisition time cost to measure the spectrum for a given bandwidth and resolution. Here, the ability to acquire dense spectral data is one of the main advantages of our approach. To illustrate this, we compare the time required to acquire a spectrum containing a peak of amplitude $1~\mu$T where the exact frequency is unknown but lies within the optimum range of either our encoding approach or conventional AC detection approaches. For our fluorescence encoding strategy, we use a representative sensitivity (Table~\ref{tab:samples}) of $10~\mu$T/$\sqrt{\text{Hz}}$, from which we estimate 100 s of data acquisition to reach a signal-to-noise level of one. However, this approach yields the entire spectrum in one measurement, enabling us to identify the peak location (and the presence of other peaks, expected or otherwise). \red{Conversely, for AC sensitivities in similar diamond samples (taken to be $1~\mu$T/$\sqrt{\text{Hz}}$ as estimated from the minimum of the Hahn Echo sensitivity trace in Figure~\ref{fig:fig2}a), we require only 1 s of data acquisition if the target frequency is known.} However, unlike the fluorescence encoding approach, acquiring a spectrum of $N$ points requires $N$ independent measurements; if 100 points are required, the time cost is equivalent for the two methods, and for $N>100$, the coherent AC sensing approach is actually slower. Thus, for spectrally congested signals, the acquisition time for the AC measurement scheme may be greater than would be suggested by the single point sensitivity. \red{We note, however, that AC sensitivities can be improved beyond our nominal $1~\mu$T/$\sqrt{\text{Hz}}$ in optimized NV center samples and measurement conditions \cite{balasubramanianUltralongSpinCoherence2009, vool2021imaging, maletinsky2012robust, palm_imaging_2022}; the time taken to acquire a spectrum with the point-scan method decreases linearly with improved sensitivity.}

\subsection{Phase-sensitive detection}
Thus far, our analysis has focused on detecting the presence of a target signal, \emph{i.e.,} measuring the power spectral density. However, measuring signals in a phase-sensitive manner is essential for many RF applications. \red{Here, we demonstrate that - with an appropriate reference or foreknowledge of the spectral structure - phase-sensitive responses can be measured, extending previous demonstrations of absolute-value spectral density detection \cite{barryOpticalMagneticDetection2016}.}

\red{In principle, the relative phase between signals in a multi-tone experiment can be extracted from the complex Fourier transform of the signal. However, this requires one continuous fluorescence time series, which is technically challenging when long averaging times are required: clock stability, periodic re-optimization of the fluorescence signal, and even computer memory constraints limit the length of a single time series. To circumvent these limitations, we introduce an approach that enables us to coherently average multiple datasets by introducing a reference signal. For an arbitrary signal of the form:
\begin{equation}
S(t) = \sum_n a_n \cos(\omega_n t)
\label{eq:bichromatic}
\end{equation}
a measurement started at an arbitrary time $t_i$ after $t = 0$, will appear to have accumulated a frequency-dependent phase $\phi_n = \omega_n t_i$. However, the phase extracted from the Fourier transform is $\phi_i \in [-\pi, \pi]$, rendering it impossible to unambiguously determine the acquired phase for \textit{all} frequencies and thus coherently average the entire complex spectra. }

\red{For a subset of frequencies, this limit can be avoided. Frequencies spaced by the greatest common divisor of $\omega_1$ and $\omega_2$ (\textit{i.e.} which satisfy $\omega_{comb} = \omega_1 + m\cdot \mathrm{gcd}(\omega_1, \omega_2)$) are subject to the same $t_0$ ambiguity, and thus are coherently averaged across multiple datasets (see \cite{supp} for further discussion). }

We demonstrate this using a test signal generated from phase modulating a 10 kHz carrier with a modulation frequency of 1 kHz and a modulation depth of $\pi/2$. We use the peaks at 10 kHz and 11 kHz (Fig.~\ref{fig:fig4}b) to phase-correct the spectrum, enabling us to coherently average 100 individual time traces and recover the correct phase-resolved spectrum at frequencies spaced by $\mathrm{gcd}(10\,\mathrm{kHz},11\,\mathrm{kHz})=1\,\mathrm{kHz}$. A requirement of this approach, however, is that the phase angles of the reference signals (and thus, $t_0$) can be reliably estimated from a single measurement, which sets a minimum amplitude for the reference signal. \redd{We use this signal to demonstrate the detection scheme because it provides an experimentally-simple phase-stable, multipeak spectrum; the approach described below is not, however, limited to detecting only phase-modulated carrier signals.}

\subsection{Spectrum detection}

\begin{figure}
\includegraphics[width=0.48\textwidth]{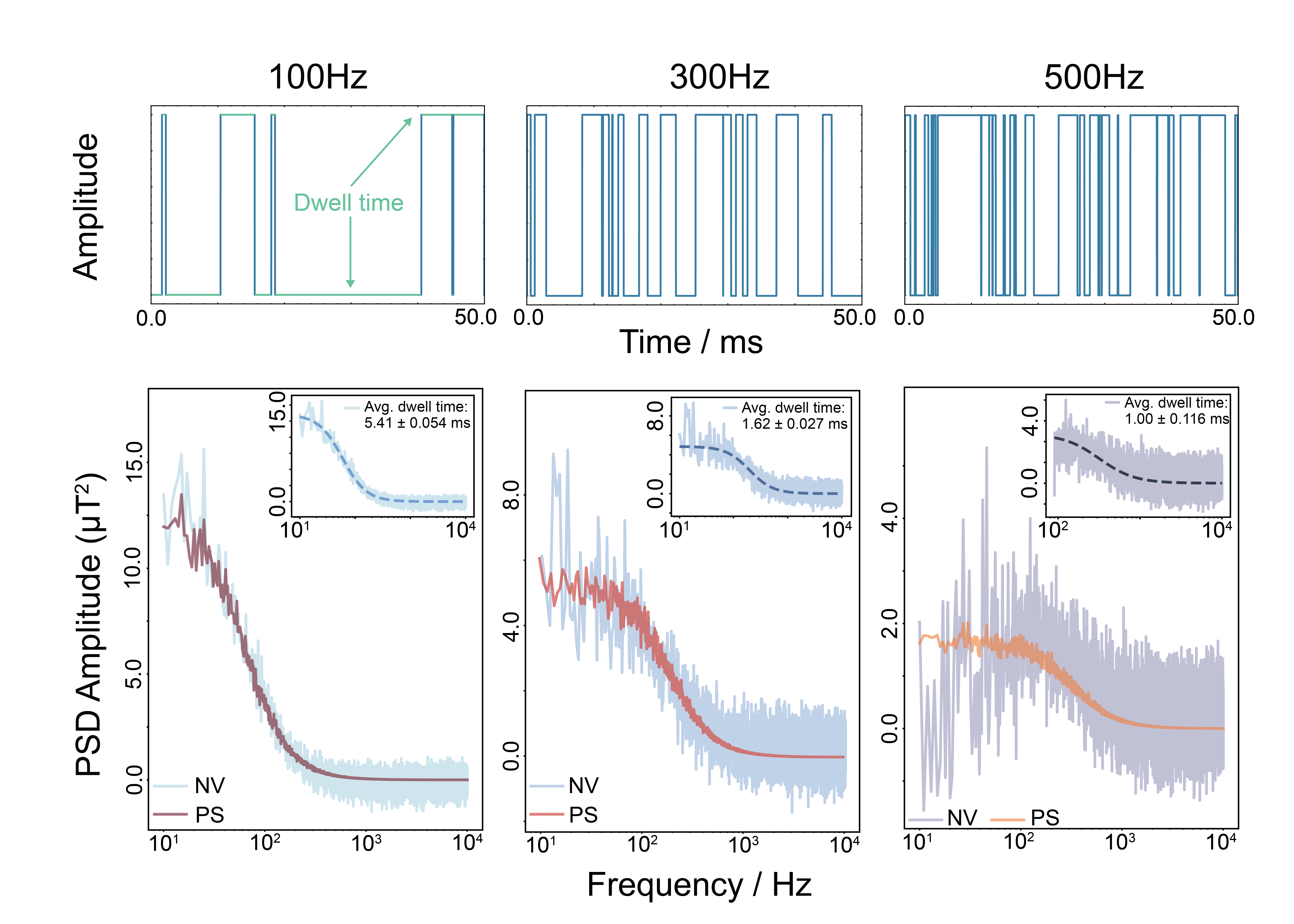}
\caption{Detection and characterization of telegraph noise with the incoherent fluorescence encoding approach. Top row shows short segments of the magnetic field trace applied to the sample. Bottom row shows the \redd{measured (labeled NV, NV-detected) and expected (labeled PS, pulse-streamer generated) PSD responses}. Bottom row insets show the measured PSD and corresponding fit from Eqn.~\ref{eq:telegraph}, yielding dwell times of (left to right) 5.41 $\pm$ 0.054 ms, 1.62 $\pm$ 0.027 ms, and 1.00 $\pm$ 0.116 ms. \red{Note that the measured data in the far right inset has been truncated to improve fit fidelity; the sloping seen in the measured data is an artifact of the filter in the audio amplifier we used to deliver the telegraph signals to our diamond sample.} }
\label{fig:fig5}
\end{figure}

Measuring spectrally congested responses is perhaps one of the biggest challenges for point-by-point measurement techniques. As a specific example, we consider two-state telegraph noise; this phenomenon is ubiquitous in nature, arising in diverse contexts such as ion channels~\cite{neherSinglechannelCurrentsRecorded1976,hamillImprovedPatchclampTechniques1981,mcmanusSamplingLogBinning1987} and enzymes~\cite{luSingleMoleculeEnzymaticDynamics1998,edmanFluctuatingEnzymeSingle1999} in biology, magnetization dynamics~\cite{hayakawaNanosecondRandomTelegraph2021,rahmanReconfigurableStochasticNeurons2024}, and charge transport in semiconductors~\cite{liRandomTelegraphNoise2018}. The dwell time for each state reveals the underlying energy landscape of the system, providing powerful new insights. Measuring these dwell times, however, is often challenging. These processes are stochastic, so time-domain averaging is not possible and requires high-fidelity real-time readout, constraining the sensitivity and/or measurement bandwidth. Instead, we consider a frequency-domain approach, since the average dwell time is encoded in the PSD response. However, this PSD is centered at DC, making it challenging to measure with coherent methods.

Here, we demonstrate that the fluorescence-encoding approach can access this rich source of information. We apply a synthetic telegraph noise signal to our sample via current delivered to a wire coil (example input traces shown in Fig.~\ref{fig:fig5} top row) and measure the resulting PSD. As shown in the bottom row of Fig.~\ref{fig:fig5}, as we decrease the average dwell time, the measured PSD broadens as expected. We fit the measured PSD traces to:
\begin{equation}
S(f) = \frac{A^2}{2T\left[\left(\frac{1}{T}\right)^2 + (\pi f)^2\right]}
\label{eq:telegraph}
\end{equation}
where $A$ is the signal amplitude, $T$ is the average dwell time, and $f$ is frequency~\cite{bothOverviewStatisticalModeling2020}.

We record the synthetic telegraph signals sent to the NV center and calculate the PSD directly; these are overlaid with the measured PSD in Fig.~\ref{fig:fig5} (bottom row) and fits to these synthetic traces are shown in Figure S9. We find that this approach is able to robustly capture the PSD of the telegraph noise with the bandwidth (and thus average dwell time; Fig. ~\ref{fig:fig5} bottom row insets) readily accessible (Table S1). \redd{Note that while we can recover the characteristic timescales of the telegraph signal, at lower frequencies slow background processes (laser fluctuations, thermal or mechanical drift) may introduce artifacts that must be mitigated.} 

\redd{Fig. \ref{fig:fig5} shows that, for a fixed time-domain amplitude (\textit{i.e.} consistent high/low levels) the amplitude of the signal in the frequency domain decreases. While the sensitivity discussion earlier is general, and still applied here, we can also consider the related question of how easily a given telegraph signal can be detected. From Eq. \ref{eq:telegraph}, we can write the maximum amplitude of our telegraph signal as $\frac{1}{2}A^2 T$. Combining this with our sensitivity allows us to describe the minimum signal which can be detected with 1 s of averaging as:
\begin{equation}
    \frac{1}{2}A^2 T = \eta ^2
\end{equation}
From this, we can see that as dwell time decreases (switching rate increases), the signal-per-frequency-bin decreases. We note, however, that this is a property of this type of signal, rather than the sensing scheme}

\red{For a calibrated sensor, the magnitude of the signal can be expressed in units of $T^2$, providing quantitative access to the power spectral density; this presents an alternative approach to direct time-domain measurements \cite{shao2016diamond}, which is particularly powerful for non-repeating, but spectrally-structured, signals.} We emphasize that these spectra are acquired without needing to know the spectral structure ahead of time, and do not rely on coherent averaging, demonstrating the utility of this approach for accessing this class of stochastic dynamics.

\section{Conclusion}

We have demonstrated a fluorescence-encoding approach for quantum sensing that expands the capability of AC magnetic field sensing. By monitoring spin-dependent fluorescence under continuous microwave excitation at the point of maximum ODMR sensitivity, we circumvent the coherence-time limitations that constrain conventional dynamical decoupling methods, enabling robust detection in the DC-MHz regime with shot-noise-limited sensitivity. This approach is technically straightforward to implement and does not require careful optimization of microwave pulse sequences.

Using this approach, we are able to transform NV centers - a model defect system - into nanoscale spectrum analyzers, simultaneously realizing broad measurement bandwidth and high spectral resolution. \redd{This capability sets the approach apart from established methods for frequency-resolved sensing. Lock-in detection, a common experimental strategy, isolates a known low-frequency modulating signal but does not reconstruct the spectrum itself. Direct time-domain detection has been used to recover transient signals, but requires either high-fidelity real-time readout or sufficient prior knowledge of the signal to apply advanced processing such as wavelet reconstruction\cite{barryOpticalMagneticDetection2016}. A more recent alternative based on Ramsey correlation measurements can improve low-frequency sensitivity\cite{zohar2026ramsey} but requires vector microwave control and a comparatively complicated filter function. While the sensitivity of our approach is limited to that of DC sensing (strictly constrained by photon detection rate, contrast, and ODMR resonance width), it combines spectrum analyzing capability with simple experimental implementation.}

Beyond the NV center platform used here, this technique is immediately applicable to the rapidly expanding ecosystem of optically-active spin qubits, including organic molecules and engineered fluorescent proteins, where short coherence times may preclude AC sensing methods. This work establishes fluorescence encoding as a complementary sensing modality that, alongside existing coherent techniques, significantly expands the operational parameter space for quantum sensing technologies.

\section{Methods}

\textit{Samples:} Two NV-containing diamond samples were used; Sample A is a nanopillar sample with individual NV centers addressable in each pillar (sourced from Qnami), while Sample B is a high-pressure high-temperature (HPHT) diamond implanted with Fe ions (Innovion, recipe listed in Table~\ref{tab:implant}) to generate a high density of vacancies~\cite{ziegler2010srim}, and subsequently annealed at 800°C for 2 hours to generate NV centers following previously reported protocols~\cite{PhysRevX.9.031052}. Additional sample details are included in \cite{supp}. 

\begin{table}
\caption{Iron implantation recipe for Sample B.}
\label{tab:implant}
\begin{ruledtabular}
\begin{tabular}{cc}
Energy (keV) & Fluence (cm$^{-2}$) \\
\hline
10 & 5$\times$10$^{12}$ \\
25 & 1$\times$10$^{13}$ \\
50 & 1$\times$10$^{13}$ \\
75 & 2$\times$10$^{13}$ \\
125 & 1.7$\times$10$^{13}$ \\
185 & 6.6$\times$10$^{13}$ \\
\end{tabular}
\end{ruledtabular}
\end{table}

\textit{Scanning confocal microscope:} All experiments were conducted using a home-built scanning confocal microscope equipped with a 532 nm laser (LaserQuantum) for excitation and an avalanche photodiode (Excelitas SPCM-AQRH-14-FC) for emission detection. In principle, the dead time of our detector ($<$50 ns) limits the detectable bandwidth using our approach (up to 10 MHz); however, as we demonstrate in the main text, this is not the main limitation on experimental bandwidth. \red{The laser is delivered to the sample through a 100$\times$ objective with a numerical aperture of 0.9 (Zeiss Objective EC Epiplan), resulting in a beam waist of $\approx300$\,nm. For all experiments, the laser power ranged from $\sim$30~$\mu$W to $\sim$300~$\mu$W; reported laser powers are measured after the objective.} For RF sensing measurements, a Time Tagger Ultra (Swabian Instruments), coupled to the avalanche photodiode, was used to timestamp each photon detection event. \red{A permanent magnet was used to apply a bias field of magnitude 75\,G; to allow efficient microwave and RF excitation in our sample geometry, the NV center probed here are oriented at $109.5^\circ$ with respect to the bias field, resulting in a smaller on-axis component of $\approx25$\,G.}

\textit{Microwave and RF delivery:} Microwaves were generated by a signal generator (Rohde \& Schwarz SMIQ 03B) and amplified (Mini-Circuits ZHL-5W-63-S+), with an amplified power of 28 dBm (Sample A) and 30 dBm (Sample B) delivered to the sample via a shorted wire $\approx 100~\mu$m from the diamond surface (Sample A) or a lithographically-defined gold stripline patterned directly on the surface (Sample B). Note, these power estimates do not account for insertion losses, etc.

RF signals (Hz-MHz) were delivered to the NV centers through one of two methods. Either the RF was combined with the microwave signal with a diplexer (Mini-Circuits ZDPLX-2150-S+), or an external wire loop ($\approx 5$ mm diameter) was used to fully decouple the microwave and RF. Narrowband and modulated RF signals were generated with a function generator (Siglent SDG6022X). The broadband noise signal was generated by digitally modulating a DC output (Red Pitaya STEMlab 125-14) with a microwave switch (Mini-Circuits ZYSWA-2-50DR+) according to predetermined, randomly-generated telegraph noise signals. This modulated signal was amplified by an audio amplifier (AIYIMA A07).

The magnetic field from the coil was calibrated externally with a Hall sensor (A1366LKT-1-T, Allegro Microsystems) and corrected for the orientation of the NV centers (here, $54.7^\circ$ from normal). More details on the calibration procedure are available in \cite{supp}.

\textit{Telegraph noise generation:} The data in Fig.~\ref{fig:fig5}b shows the measured power spectral density for synthesized telegraph noise for three average dwell times: 1 ms, 1.67 ms, and 5 ms. For each dwell time, two hundred distinct 1-second-long two-state signals were generated. Each 1-second trace was composed of dwell times sampled from a random exponential distribution according to the desired average dwell time. A single data file contained 30 seconds of measurement with the telegraph signal on (1-second traces repeated 30 times), followed by 30 seconds of measurement with the telegraph signal off. This was necessary to subtract off low-frequency ($<$10 Hz) variations and ensure the origin of the observed signals was indeed the applied noise (Figure S10). The process of 30-second two-state signal on, then 30-second two-state signal off, was repeated four times for every one of the two hundred unique signals.

\textit{Data processing:} The Time Tagger device collects the timestamps of photon events with a resolution of $\sim$2 picoseconds. The data is collected in 1-second intervals, and this collection is repeated either 30 or 60 times and saved to disk. After each file generation, the NV center position is tracked to mitigate the effects of drift. This process is repeated as required for further averaging. Saving the timestamps directly allows us to construct time series with finer or coarser resolution as required in a memory-efficient manner. (An uncompressed time-trace for a 60 s acquisition in 100 ns intervals can otherwise require many GB of disk space; writing these files can quickly become the most time-consuming part of the experiment.)

Time-tagged data were processed by binning the timestamps to generate a fluorescence time series. The width of these bins sets the upper frequency limit in our data. These time-traces were then Fourier transformed to generate the frequency-domain data shown in the main text. Unless otherwise noted, the square of the absolute value of the transform---the power spectral density---is presented.

\textit{Master equation simulations:} We use the QuTip package~\cite{johanssonQuTiP2Python2013,johanssonQuTiP2Python2013,lambertQuTiP5Quantum2025} to simulate the time-evolution of the density matrix $\rho$:
\begin{equation}
\frac{d\rho}{dt} = -i[H,\rho] + \sum_{j=1,2}\left(2L_j\rho L_j^\dagger - L_j^\dagger L_j\rho - \rho L_j^\dagger L_j\right)
\end{equation}

The Hamiltonian for the system includes both microwave (amplitude $B_1$) and RF signals (time-dependent amplitude $b(t)$). In the microwave rotating frame this yields:
\begin{equation}
H = \Delta S_z - \gamma_e b(t)S_z - \gamma_e B_1 S_x
\end{equation}
where $\Delta$ is the detuning of the system. The time-dependent RF field in principle has off-axis ($x,y$) components; the small amplitude of our signal allows us to instead focus only on the terms which commute with $S_z$.

The relaxation operators are defined as:
\begin{equation}
L_1 = \sqrt{\frac{\Gamma_1}{2}}\sigma_-, \qquad L_2 = \sqrt{\frac{\Gamma_2}{2}}\sigma_z
\end{equation}
where the Pauli spin matrices have their usual meaning. $L_1$ describes spin-lattice relaxation ($T_1$), while $L_2$ describes dephasing.

We initialize the system in a pure population state and simulate the time evolution of the density matrix. Since the fluorescence readout is a projective measurement which depends on the $S_z$ component of the NV center spin, we calculate $\langle S_z\rangle(t)$, which shows sinusoidal behavior at the same frequency as $b(t)$. To calculate the relative amplitude of the response, evaluate the peak-to-peak response after allowing enough time for the transient response to decay and a quasi-steady-state to be reached. To the calculated curves, we add a constant offset representing the noise floor of our measurement; on the log-log scale shown in Fig.~\ref{fig:fig3}, this floor is clearly visible.

\section*{Data availability} The data that support the findings of this article are openly available at \cite{zenodo}.

\begin{acknowledgments}
This material is based upon work supported by the Air Force Office of Scientific Research under award number FA9550-25-1-0098.
\end{acknowledgments}

\bibliography{references}

\end{document}